\documentclass[11pt]{article}
\usepackage{amssymb}
\usepackage{amsmath}
\usepackage{amscd}
\usepackage{latexsym}
\usepackage{bbm}

\def\a{\alpha}
\def\b{\beta}
\def\ga{\gamma}
\def\la{\lambda}

\def\eps{\epsilon}
\def\ve{\varepsilon}

\def\Si{\Sigma}
\def\p{\phi}
\def\vp{\varphi}

\def\O{\Omega}

\newcommand{\R}{\mathbb R}

\newcommand{\Gcal}{{\cal G}}
\newcommand{\Acal}{{\cal A}}

\newcommand{\Mcal}{{\cal M}}
\newcommand{\Fcal}{{\cal F}}
\newcommand{\Ncal}{{\cal N}}

\newcommand{\gfrak}{{\mathfrak g}}

\def\Tr{\textrm{Tr}}
\def\diff{\textrm{d}}
\def\pa{\mbox{$\partial$}}
\def\sfrac#1#2{{\textstyle\frac{#1}{#2}}}
\def\+{\dagger}
\def\={\ =\ }
\def\und{\qquad\textrm{and}\qquad}
\def\and{\quad\textrm{and}\quad}
\def\with{\quad\textrm{with}\quad}
\def\for{\quad\textrm{for}\quad}

\def\Id{\mathrm{Id}}

\begin{document}

\begin{titlepage}
\setcounter{page}{0}
\begin{flushright}
.
\end{flushright}

\hspace{2.0cm}

\begin{center}

{\Large\bf
Loop groups in Yang-Mills theory
}

\vspace{12mm}

{\large Alexander D. Popov}\\[8mm]

\noindent {\em
Institut f\"ur Theoretische Physik\\
Leibniz Universit\"at Hannover \\
Appelstra\ss e 2, 30167 Hannover, Germany }\\
{Email: popov@itp.uni-hannover.de
}\\[6mm]

\vspace{10mm}

\begin{abstract}
\noindent We consider the Yang-Mills equations with a matrix gauge group $G$ on the de Sitter dS$_4$,  anti-de Sitter AdS$_4$
and Minkowski $\R^{3,1}$ spaces. On all these spaces one can introduce a doubly warped metric in the form $\diff s^2 =-\diff
u^2 + f^2\diff v^2 +h^2\diff s^2_{H^2}$, where $f$ and $h$ are the functions of $u$ and $\diff s^2_{H^2}$ is the metric on the
two-dimensional hyperbolic space $H^2$. We show that in the adiabatic limit, when the metric on $H^2$ is scaled down, the
Yang-Mills equations become the sigma-model equations describing harmonic maps from a two-dimensional manifold (dS$_2$,
AdS$_2$ or $\R^{1,1}$, respectively) into the based loop group $\O G=C^\infty (S^1, G)/G$ of smooth maps from the boundary
circle  $S^1=\partial H^2$ of $H^2$ into the gauge group $G$. From this correspondence and the implicit function theorem it
follows that the moduli space of Yang-Mills theory with a gauge group $G$ in four dimensions is bijective to the moduli space
of two-dimensional sigma model with  $\Omega G$ as the target space. The sigma-model field equations can be reduced to
equations of geodesics on $\Omega G$,  solutions of which yield magnetic-type configurations of Yang-Mills fields. The group
$\Omega G$ naturally acts on their moduli space.
\end{abstract}

\end{center}
\end{titlepage}

\noindent {\bf 1. Introduction.} It is well known that the self-dual Yang-Mills equations in the Euclidean space $\R^{4,0}$
have an infinite-dimensional algebra of ``hidden symmetries'' (see~\cite{1}-\cite{6} for discovering, reviews and more
references). For the Yang-Mills potentials with value in a Lie algebra $\gfrak =\,$Lie$\,G$, where $G$ is a matrix gauge
group, among these symmetries there is the Lie algebra of the loop group $LG=C^\infty (S^1, G)$. Here we shall show that the
same group is a part of the moduli space of solutions to the Yang-Mills equations on the Lorentzian manifolds dS$_4$, AdS$_4$
and $\R^{3,1}$ of constant positive, negative and zero curvature.

We will use the {\it adiabatic limit} method  which was applied to the first-order self-dual Yang-Mills equations on the
product $\Si_1\times \Si_2$ of two Riemann surfaces in~\cite{7}. It was shown that when the metric on the Riemann surface
$\Si_2$ shrinks to a point, the Yang-Mills instantons converge to holomorphic maps from  $\Si_1$ to the moduli space of flat
connections on $\Si_2$. In~\cite{8} this limit was discussed in the framework of topological Yang-Mills theories on
$\Si_1\times \Si_2$. We will apply the adiabatic method to the second-order Yang-Mills equations on Lorentzian four-manifolds
of constant curvature and describe how to construct approximate solutions of the Yang-Mills equations. It will be shown that
these configurations become exact solutions in the adiabatic limit, when the Yang-Mills equations reduce to sigma-model
equations describing harmonic maps from two-dimensional space-time (dS$_2$, AdS$_2$ or $\R^{1,1}$) into the based loop group
$\Omega G$. For static solutions of these equations, the moduli space is the tangent space $T\Omega G$ of the based loop group
$\Omega G$. Thus, $\Omega G$ is a ``hidden symmetry group" not only of the first-order self-dual Yang-Mills equations but also
of the second-order Yang-Mills equations on Lorentzian manifolds $\R^{3,1}$, AdS$_4$ and dS$_4$.

\medskip

\noindent
{\bf 2. Metrics.} It is known that on the de Sitter dS$_4$ and anti-de Sitter AdS$_4$ spaces one can introduce (local) coordinates
such that the metrics on these spaces will be a double warped metrics of the form~\cite{9}
\begin{equation}\label{1}
\mathrm{dS}_4:\  \diff s^2_{+1} =-\diff u^2 + \cosh^2 u\, \diff v^2 +\sinh^2u\,\diff s^2_{H^2}\ ,
 \end{equation}
\begin{equation}\label{2}
\mathrm{AdS}_4:\  \diff s^2_{-1} =-\diff u^2 + \sin^2 u\, \diff v^2 +\cos^2u\,\diff s^2_{H^2}\ ,
 \end{equation}
where the first two terms are metrics on the spaces dS$_2$ and AdS$_2$, respectively. Here,
\begin{equation}\label{3}
 \diff s^2_{H^2} = \diff\chi^2 + \sinh^2\chi\, \diff\vp^2\ ,
\end{equation}
is the metric on the two-dimensional hyperbolic space $H^2$. This space has two sheets $H^2=H^2_+\cup H^2_-$ with the common
boundary $S^1$ at $\chi\to +\infty$ for $H^2_+$ and $\chi\to -\infty$ for $H^2_-$.

We introduce on the Minkowski space-time a metric similar to (\ref{1}) and (\ref{2}). In the Cartesian coordinates $x^\mu$, $\mu =0,1,2,3$,
the metric has the form
\begin{equation}\label{4}
 \diff s^2_0 = \eta_{\mu\nu}\diff x^\mu \diff x^\nu \with (\eta_{\mu\nu})=\mbox{diag}(-1, 1, 1, 1)\ .
\end{equation}
Let us introduce coordinates $u, \chi$ and $\vp$ by
\begin{equation}\label{5}
 x^0 = u\cosh\chi\ ,\quad x^1 = u\sinh\chi\cos\vp\ \and x^2 = u\sinh\chi\sin\vp\ ,
\end{equation}
and keep $x^3$ untouched. The coordinates (\ref{5}) have a range
\begin{equation}\label{6}
 u=\left((x^0)^2-(x^1)^2-(x^2))^2\right)^{1/2} > 0\ , \quad \chi\in(-\infty , +\infty)\and \vp\in [0,2\pi]\ .
\end{equation}
They cover the interior of the light cone in $\R^{2,1}$ and we denote this subset of $\R^{2,1}$ by $\R^{2,1}_+$.
The region $\R^{2,1}_-=\R^{2,1}\setminus\R^{2,1}_+$ can be covered by other choice of pseudospherical coordinates.
 In these coordinates the metric (\ref{4})  acquires the form
\begin{equation}\label{7}
 \diff s^2_0 = -\diff u^2 + (\diff x^3)^2 + u^2\,\diff s^2_{H^2}\ .
\end{equation}
From (\ref{7}) we recognize a cone over $H^2$, i.e. $\R^{2,1}_+=C(H^2)$ and we restrict ourselves to the subset
$\R^{3,1}_+=\R\times\R^{2,1}_+\subset\R^{3,1}$. For the metrics  (\ref{1}) and (\ref{2}) we also consider $u>0$, since for
$u=0$ they degenerate.

After denoting $x^3=v$, we see that the metrics (\ref{1}), (\ref{2}) and (\ref{7}) have the same form
\begin{equation}\label{8}
\diff s^2 =-\diff u^2 + f^2\diff v^2 +h^2\diff s^2_{H^2}\ ,
\end{equation}
where
\begin{equation}\nonumber
f=\cosh u\and h=\sinh u \for  \mathrm{dS}_4
\end{equation}
\begin{equation}\label{9}
f=\sin u\and h=\cos u \for  \mathrm{AdS}_4
\end{equation}
\begin{equation}\nonumber
f=1\and h= u \for  \R^{3,1}
\end{equation}
Therefore, we will consider all three spaces together by using the metric (\ref{8}), specifying $f$ and $h$ if necessary. Recall
that we work in local coordinates which cover only part of any of the considered spaces. This is enough for our purposes. For further
unification we introduce the coordinates $(y^\mu )=(y^a, y^i)=(u, v, \chi , \vp )$, where $\mu = (a,i)$ with $a, b,... =0,1$
and $i,j,...=2,3$. Then metric (\ref{8}) can be written as
\begin{equation}\label{10}
 \diff s^2=g_{\mu\nu}\diff y^\mu\diff y^\nu= g_{ab}\diff y^a\diff y^b +  g_{ij}\diff y^i\diff y^j\ ,
\end{equation}
where $g^{}_{\Si} = (g_{ab})$ is the metric on the two-dimensional space $\Si$ which is (a patch of) dS$_2$, AdS$_2$ or
$\R^{1,1}$ and $(g_{ij})= h^2(g_{ij}^{H^2})$ where $g_{H^2}=(g_{ij}^{H^2})$ is the metric on $H^2$.

Finally, as $H^2$ we will consider only the upper sheet of the two-dimensional hyperbolic space with $\chi\ge 0$ for all three
metrics (\ref{1}), (\ref{2}), (\ref{7}) and consider only $y^0=u > 0$ in (\ref{8})-(\ref{10}). All other regions of our spaces
dS$_4$, AdS$_4$, and $\R^{3,1}$ can be considered similarly.

\medskip

\noindent
{\bf 3. Yang-Mills equations.} So, we consider Yang-Mills theory on a Lorentzian 4-manifold $M$ with local coordinates $y^\mu$
and the metric given by (\ref{8})-(\ref{10}). We start with the potential $\Acal =\Acal_\mu\diff y^\mu$ with values  in the Lie algebra
$\gfrak =\,$Lie$\,G$ having scalar product defined by the trace Tr. Here $G$ is an arbitrary matrix gauge group. The field strength
$\Fcal =\diff\Acal + \Acal\wedge\Acal$ is the $\gfrak$-valued two-form:
\begin{equation}\label{11}
 \Fcal =\sfrac12\Fcal_{\mu\nu}\diff y^\mu \wedge \diff y^\nu\with \Fcal_{\mu\nu} =\partial_\mu\Acal_\nu - \partial_\nu\Acal_\mu + [\Acal_\mu , \Acal_\nu]\ .
\end{equation}
The Yang-Mills equations on $M$ with the metric given by  (\ref{8})-(\ref{10})    are
\begin{equation}\label{12}
D_\mu\Fcal^{\mu\nu}:=\frac{1}{\sqrt{|\det g|}}\,\partial_\mu(\sqrt{|\det g|}\,\Fcal^{\mu\nu}) + [\Acal_\mu ,
\Fcal^{\mu\nu}]\=0\ ,
 \end{equation}
where $g=(g_{\mu\nu})$ and indices are raised by $g^{\mu\nu}$. We have the obvious splitting
\begin{equation}\label{13}
 \Acal = \Acal_\mu\diff y^\mu = \Acal_a\diff y^a +  \Acal_i\diff y^i\ ,
\end{equation}
\begin{equation}\label{14}
 \Fcal = \sfrac12\Fcal_{ab}\diff y^a\wedge\diff y^b +  \Fcal_{ai}\diff y^a\wedge\diff y^i +  \sfrac12\Fcal_{ij}\diff y^i\wedge\diff y^j\ .
\end{equation}

\medskip

\noindent
{\bf 4. Adiabatic limit.} By using the adiabatic approach~\cite{7, 8, Sal, 10}, which is based on the ideas of~\cite{11}, we deform the metric (\ref{8})
and introduce
\begin{equation}\label{15}
\diff s^2_\ve =-\diff u^2 + f^2\diff v^2 + \ve^2h^2\diff s^2_{H^2}\ ,
\end{equation}
where $\ve$ is a real parameter. Then $|\det g_\ve |= \ve^4|\det (g_{ab})|\det (g_{ij})$ and
\begin{equation}\label{16}
 \Fcal^{ab}_\ve = g^{ac}_\ve  g^{bd}_\ve \Fcal_{cd} =\Fcal^{ab} ,\quad \Fcal^{ai}_\ve =\ve^{-2}\Fcal^{ai}\und  \Fcal^{ij}_\ve =\ve^{-4}\Fcal^{ij}\ ,
\end{equation}
where indices of $\Fcal^{\mu\nu}$ are raised by the metric  $g^{\mu\nu}$ from (\ref{10}).

To avoid the divergent term $\ve^{-2}\Tr (\Fcal_{ij}\Fcal^{ij})$ in the Lagrangian, we impose the flatness condition
\begin{equation}\label{17}
 \Fcal_{ij}=0
\end{equation}
on the Yang-Mills curvature along $H^2$ in $M$.
For the deformed metric (\ref{15}) the Yang-Mills equations have the form
\begin{equation}\label{18}
 \ve^2D_a\Fcal^{ab} + D_i\Fcal^{ib}=0\ ,
\end{equation}
\begin{equation}\label{19}
 D_a\Fcal^{aj} + \ve^{-2}D_i\Fcal^{ij}=0\ ,
\end{equation}
and in the limit $\ve\to 0$ (after choosing $\Fcal_{ij}=0$) we have
\begin{equation}\label{20}
 D_i\Fcal^{ib}=0\ ,
\end{equation}
\begin{equation}\label{21}
 D_a\Fcal^{aj}=0\ .
\end{equation}

\medskip

\noindent
{\bf 5. Flat connections.} Flat connection $\Acal_{H^2}=\Acal_i\diff y^i$ on $H^2$ (upper sheet) has a simple form
\begin{equation}\label{22}
 \Acal_{H^2}=g^{-1}\hat\diff g\with \hat\diff =\diff y^i\frac{\pa}{\pa y^i}\ ,
\end{equation}
where $g=g(y^a, y^i)$ is a smooth map from $H^2$ (for any given $y^a$) into the gauge group $G$.
We consider smooth matrix-valued functions $g$ with smooth boundary value on $S^1=\pa H^2$
and impose additional condition $g=\Id$ at $1\in S^1$ (framing of flat connection on $H^2$ \cite{Sal}). We denote by
$C^\infty_0(H^2, G)$ the space of all such $g$ in (\ref{22}). On $H^2$, as on a manifold with the boundary,
the group of gauge transformations is defined as \cite{Sal}
\begin{equation}\label{23}
 \Gcal_{H^2}= \left\{g: H^2\to G\mid {g_{|\pa H^2}} = \Id\right\}\ .
\end{equation}
Hence the solution space of the equation  $\Fcal_{H^2}=0$ is the
infinite-dimensional group $\Ncal = C^\infty_0 (H^2, G)$ and the moduli space is the based loop group \cite{12}
\begin{equation}\label{25}
 \Mcal = \Omega G=  C^\infty_0 (H^2, G)/\Gcal_{H^2}\ .
\end{equation}
Recall that $g$ and $\Acal_{H^2}$ depend on coordinates $y^a$.

\medskip

\noindent
{\bf 6. Moduli space.} On the group manifold (\ref{25}) we introduce local coordinates $\p^\a$ with
$\a =1, 2, ...$ and assume that $\Acal_\mu$'s depend on $u$ and $v$ only via
the moduli parameters $\p^\a = \p^\a (u, v)$. Then moduli of flat connections on $H^2$ define a map
\begin{equation}\label{26}
 \p : \Sigma\to\Mcal\with \p (u,v)=\{\p^\a(u,v)\}\ ,
\end{equation}
where by $\Sigma$ we denote (a patch of) dS$_2$, AdS$_2$ or $\R^{1,1}$ depending on choice of $M$ in (\ref{8})-(\ref{9}).
These maps are constrained by the equations (\ref{20}) and (\ref{21}). Since  $\Acal_{H^2}$ is a flat connection for any
$y^a\in\Sigma$, the derivatives $\pa_a\Acal_i$ have to satisfy the linearized around  $\Acal_{H^2}$ flatness condition,
i.e. $\pa_a\Acal_i$ belong to the tangent space $T_\Acal\Ncal$ of the space $\Ncal=C^\infty_0 (H^2, G)$ of flat connections on $H^2$.
Using the projection $\pi$ on the moduli space, $\pi : \Ncal\to\Mcal$, one can decompose $\pa_a\Acal_i$ into the two parts
\begin{equation}\label{27}
T_\Acal\Ncal= \pi^*T_\Acal\Mcal\oplus T_\Acal \Gcal \quad\Leftrightarrow\quad\pa_a\Acal_i=(\pa_a\p^\b)\xi_{\b i} + D_i\eps_a \ ,
\end{equation}
where $\Gcal$ is the gauge group (restricted to $H^2$ by fixing $y^a\in\Si$, $\Gcal|_{H^2}=\Gcal_{H^2}$), $\{\xi_{\a}=\xi_{\a i}\diff y^i\}$
is a local basis of tangent vectors on $T_\Acal\Mcal$ (they form the Lie algebra $\Omega\gfrak$) and $\eps_a$ are $\gfrak$-valued
gauge parameters ($D_i\eps_a$ are tangent vectors from $T_\Acal \Gcal$) which are determined by the gauge-fixing equations
\begin{equation}\label{28}
  g^{ij} D_i\xi_{\a j}=0\quad\Leftrightarrow\quad g^{ij} D_i\pa_a\Acal_j= g^{ij} D_iD_j\eps_a\ .
\end{equation}
In fact, since $\p^\a$ depend on $y^a\in\Si$, we have $\Ncal=\Ncal(y^a)$, $\Gcal=\Gcal_{H^2}(y^a)$ and $\Mcal =\Ncal /\Gcal=\Mcal(y^a)$.

Recall that $\Acal_i$ are fixed by (\ref{22}) and  $\Acal_a$ are yet free. For the mixed components
of the field strength we have
\begin{equation}\label{29}
 \Fcal_{ai}=\pa_a\Acal_i - D_i\Acal_a = (\pa_a\phi^\b)\xi_{\b i}- D_i(\Acal_a-\eps_a)\ .
\end{equation}
It is natural to choose $\Acal_a=\eps_a$ \cite{11, 13} and obtain
\begin{equation}\label{30}
 \Fcal_{ai}=(\pa_a\phi^\b)\xi_{\b i}=\pi_*\pa_a\Acal_i \in T_\Acal\Mcal\ .
\end{equation}
On the other hand, since $\Acal_i(\p^\a , y^j)$ depends on $y^a$ only via $\p^\a$, we have
\begin{equation}\label{31}
 \pa_a\Acal_i=\frac{ \pa\Acal_i}{ \pa\phi^\b} \frac{ \pa\phi^\b}{ \pa y^a}\quad\stackrel{\mathrm{(27)}}{\Longrightarrow}\quad \eps_a=\Acal_a=
 (\pa_a\phi^\b)\,\eps_\b
\end{equation}
with the gauge parameters $\eps_\a$ defined by (\ref{28}) via the equations
\begin{equation}\label{32}
g^{ij} D_i D_j\eps_\a=g^{ij} D_i\frac{ \pa\Acal_j}{ \pa\phi^\a}\ .
\end{equation}
Thus, if we know $\p^\a (u,v)$ then we can construct
\begin{equation}\label{33}
 (\Acal_\mu )= (\Acal_a, \Acal_i)=\left( (\pa_a\phi^\b)\,\eps_\b\ ,\ g^{-1}(\p^\b , y^j)\pa_i g(\p^\b , y^k)\right)
\end{equation}
which should solve the equations (\ref{20}) and (\ref{21}).

\medskip

\noindent
{\bf 7. Harmonic maps.} Substituting (\ref{30}) into (\ref{20}), one can see that these equations
are resolved due to (\ref{28}), (\ref{31}) and (\ref{32}). On the other hand, substitution of (\ref{30}) and (\ref{31}) into (\ref{21})
gives us the equations
\begin{equation}\label{34}
\frac{1}{\sqrt{|\det g^{}_{\Si}|}}\ \pa_a\left(\sqrt{|\det g^{}_{\Si}|}\ g^{ab}\pa_b\p^\b\right)g^{ij}_{H^2}\xi_{\b j}+
g^{ab}g^{ij}_{H^2}(D_a\xi_{\b j})\pa_b\p^\b= 0\ ,
\end{equation}
where $g^{}_{\Si} = (g_{ab})$ is the metric on $\Si$ and $g_{H^2}=(g_{ij}^{H^2})$ is the metric on $H^2$. Before proceeding
further we introduce a metric ${\mathbb G}=(G_{\a\b})$ on the moduli space $\Mcal =\Omega G$ of flat connections on $H^2$ as
\begin{equation}\label{35}
 G_{\a\b}\equiv \langle \xi_\a , \xi_\b\rangle = - \int_{H^2} d\, \mbox{vol} \ g^{ij}_{H^2}\Tr(\xi_{\a i} \xi_{\b j})\ ,
\end{equation}
where the integral is taken over $H^2=H^2_+$. Multiplying (\ref{34}) by $\xi_{\a i}$ and integrating over $H^2$ (projection on
the moduli space) , we obtain
\begin{equation}\nonumber
 \frac{1}{\sqrt{|\det g^{}_{\Si}|}}\,\pa_a\left(\sqrt{|\det g^{}_{\Si}|}\ g^{ab}\pa_b\p^\b\right)\langle \xi_\a , \xi_\b\rangle +  g^{ab}\langle \xi_\a ,
 D_a\xi_\b\rangle\pa_b\p^\b
\end{equation}
\begin{equation}\label{36}
= \frac{1}{\sqrt{|\det g^{}_{\Si}|}}\,\pa_a\left(\sqrt{|\det g^{}_{\Si}|}\ g^{ab}\pa_b\p^\b\right)G_{\a\b} + \langle \xi_\a ,
\nabla_\gamma\xi_\b\rangle g^{ab}\pa_a\p^\gamma \pa_b\p^\b
\end{equation}
\begin{equation}\nonumber
 =G_{\a\sigma}\left\{\frac{1}{\sqrt{|\det g^{}_{\Si}|}}\,\pa_a\left(\sqrt{|\det g^{}_{\Si}|}\ g^{ab}\pa_b\p^\sigma\right) +
 \Gamma^\sigma_{\b\ga}g^{ab}\pa_a\p^\b\pa_b\p^\ga\right\} =0\ ,
\end{equation}
where
\begin{equation}\label{37}
 \Gamma^\sigma_{\b\ga}=\sfrac12\, G^{\sigma\la}\left(\pa_\ga \, G_{\b\la} +
 \pa_\b G_{\ga\la} - \pa_\lambda G_{\b\ga}\right)\with \pa_\ga :=\frac{\partial}{\partial \phi^\ga}\ ,
\end{equation}
are the Christoffel symbols and $\nabla_\ga$ are the corresponding covariant derivatives on the moduli space $\Omega G$ of
flat connections on $H^2$.

The equations
\begin{equation}\label{38}
\frac{1}{\sqrt{|\det g^{}_{\Si}|}}\,\pa_a\left(\sqrt{|\det g^{}_{\Si}|}\ g^{ab}\pa_b\p^\a\right) +
\Gamma^\a_{\b\ga}g^{ab}\pa_a\p^\b\pa_b\p^\ga =0
\end{equation}
are the Euler-Lagrange equations for the effective action
\begin{equation}\label{39}
 S_{\rm eff}=\int_{\Si}\diff y^1\diff y^2 \sqrt{|\det g^{}_{\Si}|}\    \, g^{ab}G_{\a\b}\pa_a\p^\a \pa_b\p^\b
\end{equation}
of the Yang-Mills theory on $M$ which appears from the term Tr$(\Fcal_{aj}\Fcal^{aj})$ in the initial Yang-Mills Lagrangian in
the adiabatic limit $\ve\to 0$. The equations (\ref{38}) are the standard sigma-model equations defining harmonic maps from
$\Sigma\, ( =\,$dS$_2$, AdS$_2$ or $\R^{1,1}$) into the based loop group $\Omega G$ parameterized (locally) by coordinates
$\p^\a$. Note that these equations are integrable and their  solutions can be constructed similar to those for
finite-dimensional target Lie groups (see e.g. \cite{14a}). If $\p^\a$ do not depend on $u$ then equations (\ref{38}) reduce
to the geodesic equations on the loop group $\O G$ and give static configurations of Yang-Mills fields. This case of magnetic
configurations can be considered as supplementing \cite{LP}, where only electric components  of adiabatic Yang-Mills fields
were nonvanishing. Note that any geodesic on $\Omega G$ is parametrized by the initial point $\phi_0\in \Omega G$ and by the
velocity $\phi_0^\prime\in T_0\Omega G$. Therefore, the moduli space of solutions (\ref{33}) (with $\pa_u\phi =0$ and
$\pa_v\phi =: \phi^\prime$) can be identified with the tangent bundle $T\Omega G$ of $\Omega G$. The based loop group $\Omega
G$ naturally acts on $T\Omega G$ which can be identified with the semi-direct product $\Omega G\ltimes \gfrak$ of $\Omega G$
and the Lie algebra $\gfrak = \,$Lie$G$.

\medskip

\noindent {\bf 8. Concluding remarks.} In conclusion we recall that in the Euclidean case Atiyah has shown \cite{14} that the
moduli space of instantons over $\R^{4, 0}$ is bijective to the moduli space of {\it holomorphic} maps from $S^2$ to $\Omega
G$. There is a conjecture (see e.g. \cite{15}) that the moduli space of solutions of the second-order Yang-Mills equations  on
$\R^{4, 0}$ is bijective to the moduli space of {\it harmonic} maps from $S^2$ to $\Omega G$. Our consideration in this paper
can be repeated for the Euclidean space. In  \cite{16} it was observed that $\R^4\cup \{\infty\}\setminus S^1=S^4\setminus
S^1$ is conformally diffeomorphic to the product manifold $S^2\times H^2$. Considering $M=S^2\times H^2$ and literally
repeating our calculations for this Euclidean manifold we will arrive to the equations (\ref{38})  with $\Si =S^2$. These
equations will define harmonic maps from $S^2$ into the based loop group $\Omega G$. Furthermore, from the implicit function
theorem it follows that near every solution $\p$ of  (\ref{38}) with $\Si =S^2$ (and the corresponding solution
$\Acal^{\ve=0}$ of the Yang-Mills equations) there exists a solution $\Acal^{\ve >0}$ of the Yang-Mills equations  on $M$ for
$\ve$ sufficiently small (cf. with the instanton case \cite{7, Sal, 16}). In other words, solutions of (\ref{38}) with $\Si
=S^2$ approximate solutions of the Yang-Mills equations on $M=S^2\times H^2$ (and on $\R^{4,0}$ after some maps \cite{Sal, 16}
from $S^2\times H^2$ to $\R^{4,0}$) and one can conjecture that the moduli spaces for $\Acal^{\ve=0}$ and $\Acal^{\ve >0}$ are
bijective. Our consideration shows that the same conjecture is valid for Lorentzian signature, i.e. it is reasonable to expect that the
Yang-Mills model  on Minkowski space $\R^{3,1}$ with a gauge group $G$ is equivalent to the sigma model on $\R^{1,1}$ whose target
space is the based loop group  $\O G$.

\bigskip

\noindent
{\bf Acknowledgements}

\medskip

\noindent
I would like to thank Olaf Lechtenfeld for valuable remarks.
This work was partially supported by the Deutsche Forschungsgemeinschaft grant LE 838/13.


\end{document}